\documentclass[runningheads]{llncs}
\usepackage[T1]{fontenc}
\usepackage{graphicx}
\usepackage{color}
\usepackage{biblatex}
\usepackage{hyperref}
\usepackage{todonotes}

\title{Beatnik: A Novel Global Communication Mini-Application}
\author{Jason R.~Stewart\inst{1} \and Patrick G.~Bridges\inst{2}\orcidID{0000-0003-4801-0390}}
\institute{Department of Computer Science, University of New Mexico\\
\email{jastewart@unm.edu}
\and
Center for Advanced Research Computing, University of New Mexico\\
\email{patrickb@unm.edu}}

\date{June 2024}

\begin{document}

\maketitle

\begin{abstract}
Beatnik is a novel open source mini-application that exercises the complex communication patterns often found in production codes but rarely found in benchmarks or mini-applications. It simulates 3D Raleigh-Taylor instabilities based on Pandya and Shkoller’s Z-Model formulation using the Cabana performance portability framework. This paper presents both the high-level design and important implementation details about Beatnik along with four benchmark setups for evaluating different aspects of HPC communication system performance. Evaluation results demonstrate both Beatnik's scalability on modern accelerator-based systems using weak and strong scaling tests up to 1024 GPUs, along with Beatnik's ability to expose communication challenges in modern systems and solver libraries.

\end{abstract}

\section{Introduction}
\label{sec:intro}
Mini-applications that reproduce and exercise key features of important numerical methods and production applications have become an essential part of the toolbox for designing and evaluating high-performance computing (HPC) systems. 
Existing mini-applications with significant communication components, however, generally exercise only simple communication patterns and primitives, most frequently either:
\begin{itemize}
    \item Neighbor exchanges for regular meshes or particle distributions~\cite{heroux_mantevo_nodate}, irregular or adaptive meshes~\cite{franko_cfd_2015, henning_ume_nodate, robey_clamr_nodate}, or sparse matricies~\cite{yang_amg2023_nodate}; or 
    \item Simple global collectives such as \verb|MPI_Allreduce| for computing dot products~\cite{dongarra_new_2016}, \verb|MPI_Allgather| for neighbor identification~\cite{yang_amg2023_nodate,robey_clamr_nodate}, or \verb|MPI_Alltoall| for matrix transposition.
\end{itemize}

Few if any HPC mini-applications, however, exercise the varied, complex communication patterns common in production applications. This is, we hypothesize, because both the numerical methods that require these complex communication patterns are inherently complex and beyond the scope of what is normally considered appropriate for a mini-application. The result, however, is that the performance of important communication patterns (e.g. those resulting the from remeshing, particle/mesh migration in ALE codes, or tree sweeps in fast multipole methods) are generally not considered when optimizing MPI communication operations.

This paper presents Beatnik, a novel open source~\cite{bridges_beatnik_nodate} mini-application designed to exercise the complex global communication patterns found in real production codes on modern GPU-accelerated systems. Beatnik does so in a modestly-sized code base (less than 5000 source lines of C++ code) by leveraging recent numerical method innovations and modern performance portability libraries~\cite{slattery_cabana_2022, carter_edwards_kokkos_2014, krzhizhanovskaya_heffte_2020, lebrun-grandie_arborx_2021, noauthor_silo_nodate}. Specifically, this paper describes the following contributions:
\begin{itemize}
\item The overall goals and architecture of Beatnik, a mini-application that simulates Raleigh-Taylor instabilities in 3D based on Pandya and Shkoller's Z-Model formulation~\cite{pandya20233d}. The chosen numerical method and benchmark architecture support multiple solution approaches that can exercise a wide range of important global communication patterns not found in many other mini-applications (Section~\ref{sec:architecture});
\item The initial Beatnik software design and implementation built using the Cabana performance portability framework and associated libraries, including high-, medium-, and low-order solver implementations that leverage large-scale Fast Fourier Transforms (FFTs) and multi-dimensional data redistributions (Section~\ref{sec:implementation}); 
\item The identification of Beatnik test cases that exercise different elements of MPI communication such as network bandwidth, network latency, and load imbalance
(Section~\ref{sec:test-cases}); and,
\item The demonstration of the  scalability of the Beatnik implementation on these test cases and its ability to exercise communication performance on modern machines up to 1024 GPUs (Section~\ref{sec:benchmark-evaulation}).
\end{itemize}
Following the presentation of these contributions, the paper discusses directions for future work
 (Section~\ref{sec:future}), and concludes (Section~\ref{sec:conclusion}).




\section{Benchmark Goals and Architecture}
\label{sec:architecture}

The primary goal of Beatnik is to provide a communication benchmark that exercises numerical solvers with irregular and/or global communication patterns while still being relatively straightforward to understand and implement. As a first step, we first sought to identify a numerical method that is scalable, straightforward to understand, straightforward to implement on modern architectures, and can leverage a range of numerical algorithms which require global communication. 

We settled on Pandya and Shkoller's so-called ``Z-Model''~\cite{pandya20233d} for simulating fluid interface instabilities as a numerical method that met these requirements. In particular, the Z-Model is an ideal method for our global communication benchmarking purposes because it includes low-, medium-, and high-order methods that exercise a range of important numerical algorithms with global communication requirements. Specifically, the low-order solver uses Fast Fourier Transforms (FFTs)---and by extension all-to-all communication---to approximate the Birkhoff-Rott velocity integral. In contrast, the high-order model requires solving the Birkhoff-Rott integral directly using any of a number of far-field force solution methods, for example the fast multipole method (FFM) or particle–particle
–particle–mesh (P3M), both of which require comnplex global communication. Finally, the medium-order solver \emph{couples} the FFT solver and the far-field force solver, resulting multiple complex coupled communication patterns and data redistributions. 

The overall architecture of Beatnik consists of the following high-level modules:
\begin{description}
    \item[SurfaceMesh:] This module implements the distributed 2D mesh that represents the interface between two fluids; each point on this mesh stores the x/y/z coordinate of the point and two vorticity components. The current Z-Model formulation is based on an open regular rectangular mesh and this formulation is the basis of Beatnik; an unstructured Z-model formulation that could more easily support closed surfaces is planned for development but has not yet been finalized. 
    \item[BRSolver:] This module calculates the Birkhoff-Rott velocity integral for points on the \texttt{SurfaceMesh} from the vorticity and position of every point on the mesh. This requires using an appropriate far-field force calculation method, either internally implemented or provided by an external solver. This module, used only by high- and medium- order Beatnik solves, is the main source of irregular global communication in the benchmark;
    \item[ZModel:] This module calculates the derivatives of position and vorticity on points on the \texttt{SurfaceMesh} using either the low-, medium-, or high- order method. To do so, it uses an external FFT solver, the \texttt{BRSolver} module, and laplacians and surface normals calculated from the \texttt{SurfaceMesh} as determined by the requested order of the ZModel solve;
    \item[TimeIntegration:] This module uses derivatives calculated using the \texttt{ZModel} module to perform time integration of points on \texttt{SurfaceMesh}, resulting in the update of the \texttt{SurfaceMesh} for the next timestep.
\end{description}

The fundamental decomposition of data resulting in MPI communication in Beatnik is of points on the \texttt{SurfaceMesh}. A 2D block decompostion of this mesh is most natural decause every derivative calculation in the ZModel module requires calculating surface normals and laplacians, distributed FFT solvers generally expect block-decomposed data, and and the current Z-Model formulation is based on a regular 2D mesh. However, other data decompositions are possible, for example ones based on the x/y/z location of the mesh point to optimize far-field force calculations.

\section{Implementation}
\label{sec:implementation}

Starting from the high-level architecture described in the previous section, we implemented Beatnik in C++ using the Cabana particle/grid library~\cite{slattery_cabana_2022}. Cabana uses the Kokkos performance portability ecosystem~\cite{carter_edwards_kokkos_2014} to support portable execution on a wide range of modern hardware architectures, and provides robust distributed mesh abstractions as well as interfaces to multiple external solvers valuable for implementing Beatnik. Key among the external libraries to which Cabana provided access were the heFFTe GPU-accelerated fast Fourier transform library~\cite{krzhizhanovskaya_heffte_2020}, the ArborX library for geometric neighbor search~\cite{lebrun-grandie_arborx_2021}, and the Silo I/O library~\cite{noauthor_silo_nodate}. The availability of robust mesh data structures and key solver interfaces in Cabana dramatically reduced the size and complexity of the Beatnik code base.

Beatnik's current C++ implementation consists of approximately 3800 source lines of code grouped into two set of modules: core modules which implement fundamental data structures and algorithms common to all solution methods, and Birchoff-Rott (BR) solver modules which implement different approaches to estimating the far-field forces required by the Z-Model's high-order and medium-order solvers. We detail these modules and their communication behavior in the remainder of this section.

\subsection{Core Modules}
The core module structure of Beatnik was taken from the Cabana ExaMPM benchmark~\cite{noauthor_exampm_nodate}. In particular, Beatnik's core code is arranged as a C++ header-only library for use by driver programs that specify specific problems to solve. The following C++ classes providing core Z-Model solution functionality:
\begin{description}
    \item[Solver] initializes and invokes other classes based on parameters passed by the driver program and runs the simulations for the specified number of timesteps.
    \item[SurfaceMesh] holds the Cabana structured grid object that encompasses the 2D domain of the interface between the two fluids. This mesh is decomposed between MPI processes using a regular 2D block decomposition. Beatnik uses Cabana's halo exchange primitives on the \texttt{SurfaceMesh} to perform two-node-deep stencils for calculating surface normals, finite differences, and Laplacians along the surface.
    \item[BoundaryCondition] applies periodic or non-periodic boundary conditions to the nodes in the \texttt{SurfaceMesh}; most boundary condition handling is provided by the underlying Cabana mesh object, but the \texttt{BoundaryCondition} explicitly corrects x/y/z coordinates in ghost cells  for periodic boundary conditions and extrapolates position and vorticity into boundary cells for non-periodic boundary conditions. No additional communication is required for these calculations. 
    \item[ProblemManager] stores the mesh state shared between multiple classes and maintained across multiple processes, primarily the position and vorticity values on the \texttt{SurfaceMesh} for the current time step. It also provides interfaces to the Cabana halo exchange for both \texttt{SurfaceMesh} state stored by the \texttt{ProblemManager} as well for use by any solvers that store their own \texttt{SurfaceMesh} objects for internal state. 
    \item[ZModel] computes derivatives of interface position and velocity using heFFTe and BR solver modules as necessary for different model orders. The model order used by an instance of the \texttt{ZModel} class is determined by a template type tag (\texttt{Order::Low, Order::Medium, Order::High}) that chooses the appropriate specialized derivative calculation method. The \texttt{ZModel} class does \emph{not} directly communicate with other MPI processes but instead invokes other classes which carry out communication specific to the relevant numerical method, including \texttt{SurfaceMesh} regular halos, heFFTe all-to-all communication, and any needed global communication by BRSolver modules.
    \item[TimeIntegrator] solves for the position and vorticity of each point on the SurfaceMesh using third-order Runge-Kutta interpolation. Because this is a third-order method, it calculates three derivatives and hence invokes the \texttt{ZModel} object three times per timestep.
    \item[SiloWriter] uses the Silo library to write surface mesh data for visualization.
\end{description}

\subsection{Birchoff-Rott Solvers}

The Birchoff-Rott solvers for medium and high-order models are generally the most complex and computationally and communication-intensive portions of Beatnik. Because of this, the Beatnik implementation supports multiple Birchoff-Rott solver classes which can be selected by the invoking driver code and ZModel solver with appropriate template arguments. Each Birchoff-Rott solver computes the far-field forces between all points on the \texttt{SurfaceMesh}, and a wide range of numerical methods have been developed for scalability approximating the far-field forces, all of which require complex communication and computation.

Beatnik currently implements two relatively simple strategies for calculating Birchoff-Rott integrals, with implementation of additional strategies planned for future work as described in Section~\ref{sec:future}. 

\subsubsection{Exact Integral Calculation}
The \texttt{ExactBRSolver} class calculates the exact Birkhoff-Rott velocity integral using a brute force all-pairs approach. We include an exact solver for the Birchoff-Rott integral despite its high computational complexity ($O(n^2)$ in the number of \texttt{SurfaceMesh} points) to enable evaluation of the accuracy/performance tradeoffs of approximate Birchoff-Rott solvers. \texttt{ExactBRSolver} uses a standard ring-pass communication algorithm to circulate \texttt{SurfaceMesh} points between processes in parallel with computing forces between mesh points. This results in regular communication patterns, but the \texttt{ExactBRSolver} is compute bound, not communication bound.

\subsubsection{Cutoff-based Approximate Integral Calculation}

The \texttt{CutoffBRSolver} is the first (and currently only) scalable far-field force solver currently implemented in Beatnik. It approximately solves the Birchoff-Rott integral by only considering the points within a specified 3D distance of each point on the \texttt{SurfaceMesh} being evaluated. The cutoff distance determines the tradeoff between accuracy and performance in this method; small cutoff distances result in better scalability at the expense of numerical inaccuracy, while high cutoff distances increase numerical accuracy but reduce performance by increasing the number of pairs of forces thtat must be calculated. Note that unlike more sophisticated far-field solution methods like the fast multipole method, this method does not allow directly setting numerical accuracy tolerances; instead the programmer choosing the cutoff distance must set the cutoff distance most appropriate for the problem being simulated.

The cutoff solver produces dynamic and irregular communication patterns because its computation is based on the \emph{3D spatial location} of \texttt{SurfaceMesh} points which are (1) decomposed between MPI processes based on the \emph{2D surface index} of the point and (2) which change as the interface surface evolves. Specifically, the cutoff solver works by, for \emph{each derivative calculation}:
\begin{enumerate}
\item Migrating each 2D \texttt{SurfaceMesh} node into a 3D spatial mesh data structure decomposed by x/y/z location.
\item Haloing points between spatial mesh blocks on differnt processors so that each process has copies of all points within the cutoff distance of each point in the spatial mesh that a processor owns.
\item Using the ArborX solver to compute local neighbor lists for each owned interface point.
\item Calculating the force on each owned interface point by each point on the neighbor list.
\item Migrating each interface point back to its original 2D \texttt{SurfaceMesh} decomposition for use by the calling \texttt{ZModel} module (which works in the 2D surface decomposition).
\end{enumerate}

To do so, \texttt{CutoffSolver} uses two supporting C++ classes:
\begin{description}
    \item[SpatialMesh] holds the Cabana structured grid object that encompasses the 3D domain of the interface between the two fluids. We currently use a 2D x/y block decomposition of the 3D space to mirror the initial distribution of 2D surface points and reduce load imbalance, though load imbalance between these blocks can develop as surface point spatial locations change. 
    \item[HaloComm] handles the migration and haloing of nodes between the surface and spatial mesh based on their x/y/z positions. This redistribution code originated from the CabanaPD benchmark~\cite{reeve_ornlcabanapd_2022} with slight modifications to fit the design of Beatnik.    
\end{description}

\section{Mini-Application Driver Problem and Input Decks}
\label{sec:test-cases}

Building on the solver implementation described in the previous section, we implemented a driver for the well-known rocket rig problem as the primary numerical problem to use for Beatnik benchmarking. This problem simulates the Rayleigh-Taylor instabilities that develop on the interface between two fluids of different densities due to their acceleration in the Z direction. The rocket rig driver program, comprising approximately 700 source lines of code, initializes and solves variants of this problem with configurable initial conditions, boundary conditions, and simulation parameters (e.g., cutoff distance). 

\begin{figure}[ht]
\centering
\includegraphics[width=0.9\textwidth]{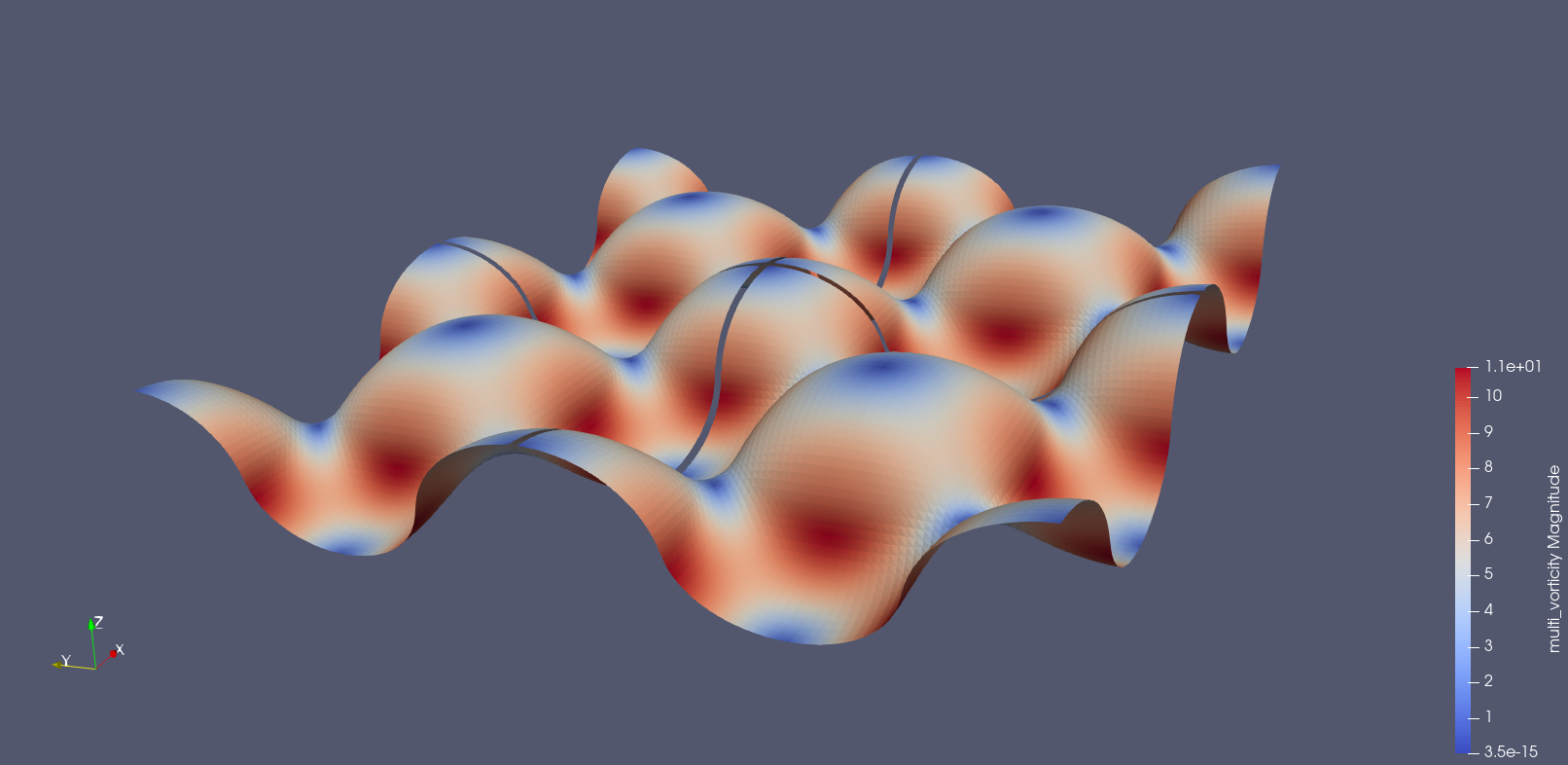}
\caption{4-GPU multi-mode test case output using low-order solver at timestep 20 with mesh points colored by vorticity magnitude}
\label{fig:low-strong-io}
\end{figure}

\begin{figure}[ht]
\centering
\includegraphics[width=0.9\textwidth]{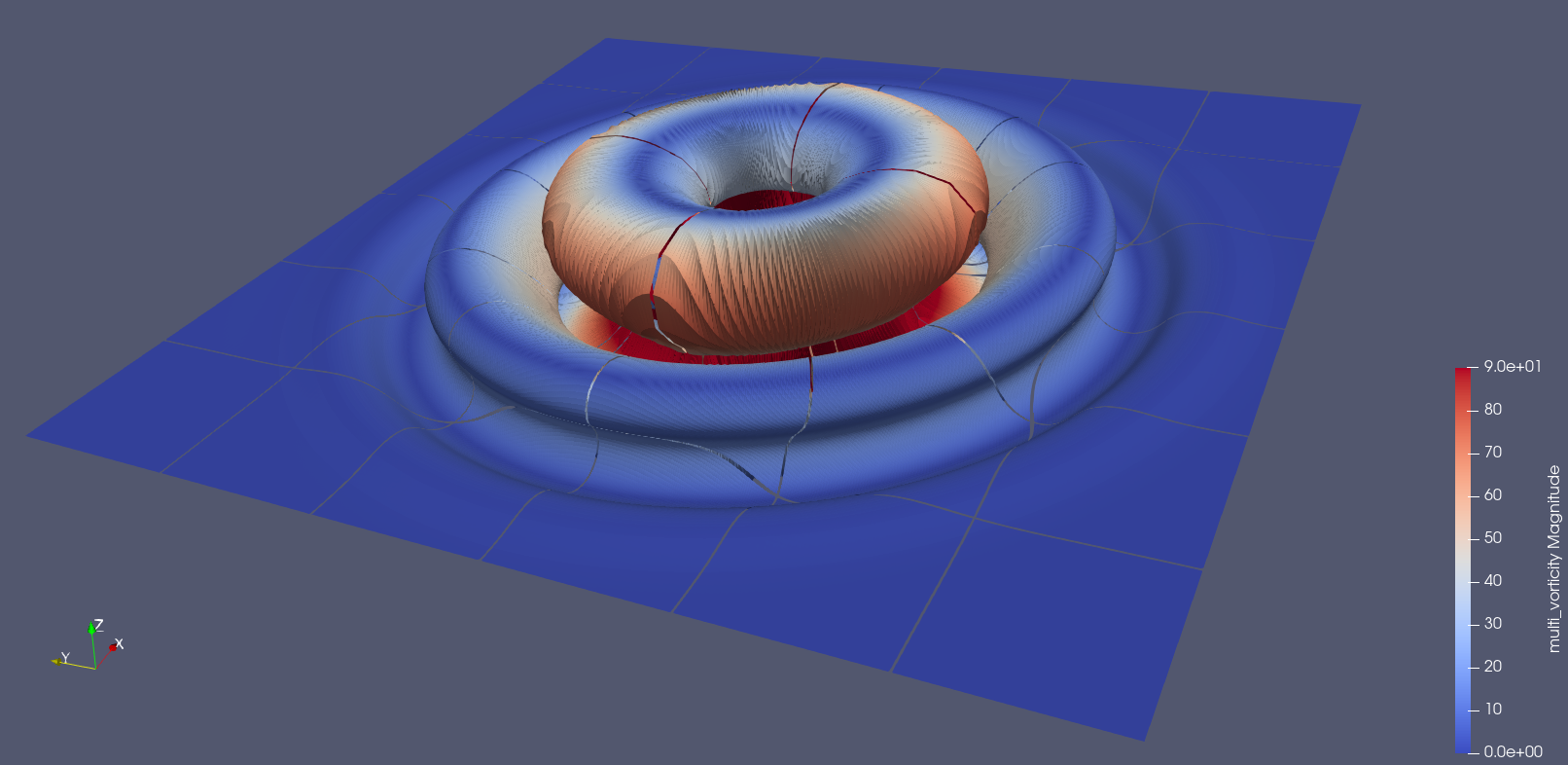}
\caption{36-GPU single-mode test case output using the cutoff solver at timestep 340 with mesh points colored by vorticity magnitude}
\label{fig:high-strong-io}
\end{figure}

For benchmarking MPI communication, we identified two specific rocket rig test cases: a multi-mode periodic test case shown in Figure~\ref{fig:low-strong-io} and a single-mode non-periodic test case shown in Figure~\ref{fig:high-strong-io}. The multi-mode test case results in relatively even particle distributions in the X/Y direction, limiting load imbalance. It is also more amenable to low-order solutions that require FFTs due to its periodic nature; Beatnik does not currently support low- and medium-order solutions of non-periodic boundaries because of its reliance on periodic FFT solvers. In contrast, the non-periodic single-mode test cases results in significant load imbalance due to surface rollup but requires a high-order solver to fully resolve these effects. It is important to note that the single-mode test case is also a useful test case for evaluating the accuracy of Birchoff-Rott solvers; the outer rollup show in Figure~\ref{fig:high-strong-io} will not develop without inclusion of distant far-field surface points. 

Based on these two test cases, we identified the following benchmark test cases to stress different MPI communication aspects:
\begin{itemize}
    \item A \textbf{multi-mode low-order weak scaling} test case that stresses 
    network bandwidth through all-to-all communication of the growing global \texttt{SurfaceMesh} by the FFT solver.
    \item A \textbf{multi-mode low-order strong scaling} test case that stresses network latency through all-to-all communication of the fixed size \texttt{SurfaceMesh} on increasing numbers of process by the FFT solver.
    \item A \textbf{multi-mode high-order weak scaling} test case using the cutoff solver to test the general scalability of MPI communication with increasing process counts.
    \item A \textbf{single-mode high-order strong scaling} test case using the cutoff solver to create load imbalances in communication and computation, and to test MPI communication performance with dynamic and irregular communication patterns.
\end{itemize}
This wide range of potential test cases highlights the flexibility of the Beatnik mini-application design to exercise a wide range of global communication features.

\section{Benchmark Evaluation}
\label{sec:benchmark-evaulation}
To demonstrate Beatnik efficacy in exercising communication system, we present scaling studies for each of the test cases identified in Section~\ref{sec:test-cases}. We also demonstrate the use of the Beatnik low-order weak scaling test case to evaluate the performance of different heFFTe communication configurations. 

\subsection{Experimental Setup}
\label{subsec:setup}
All tests described in this section were run on the Lassen system at Lawrence Livermore National Laboratories. Lassen is an IBM Power9 system with 40 usable cores per node, 4 V100 16GB GPUs per node, and an Infiniband EDR interconnect. All test cases were run with one MPI process and one Power9 core per GPU and varying the number of processes and GPUs from 4 to 1024 for each test case. 

Beatnik on Lassen was compiled using Spack (a Spack configuration for compiling Beatnik is available in the standard Spack repository). On Lassen, this results in linking against the current master branch of Cabana, Kokkos 4.2.0, heFFTe 2.4.0, ArborX 1.5, Silo 4.11.1, CUDA 11.1.1, and the rolling release of IBM Spectrum MPI. Spectrum MPI was configured to use GPU-aware MPI, preventing us from using newer CUDA versions due to a known bug in Spectrum MPI.

For both low-order and high-order tests, we set a base problem size to use all of the memory of each GPU and then either weak or strong scale this problem size to the appropriate number of processes and GPUs for the test case. For low-order problems, this results in a (-19, -19, -19) to (19, 19, 19) spatial domain and 4864 mesh points in the x and y dimensions (roughly 24M mesh points). For high-order cutoff solver test cases, this results in a (-3, -3, -3) to (3, 3, 3) spatial domain. For the the multi-mode high-order weak scaling test we use a single GPU base mesh size of 768x768 (roughly 600k points) and a cutoff distance of 0.2. For the single-mode high-order strong scaling test, we use a mesh size of 512x512 (roughly 250k points) and a cutoff distance of 0.5; tests with smaller cutoff distances resulted in significant numerical inaccuracy. 

Note that the high-order solver is more memory intensive than the low-order solver because the high-order solver must construct neighbor lists for each point to do derivative calculations. Similarly, the single mode high-order test case is more memory intensive that the multi-mode high-order test case because of longer neighbor lists on some processes due to load imbalance.

\subsection{Low-Order Solver Weak and Strong Scaling}
To test Beatnik's ability to exercise communication bandwidth and latency limits on modern systems, we weak and strong scaled the low-order solver, whose performance is limited by FFT performance, from 4 to 1024 GPUs on Lassen. 
\label{subsec:low-scaling}
\begin{figure}[ht!]
\centering
\includegraphics[width=0.7\textwidth]{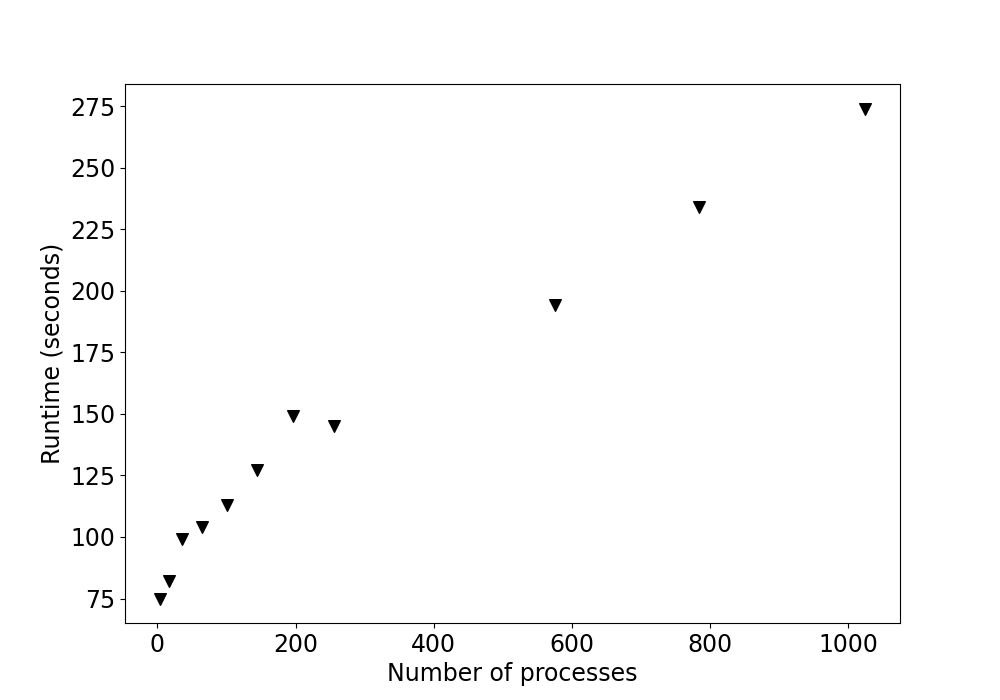}
\caption{Low-Order Weak Scaling of Beatnik on Lassen}
\label{fig:low-weak-scale}
\end{figure}
\begin{figure}[ht!]
\centering
\includegraphics[width=0.7\textwidth]{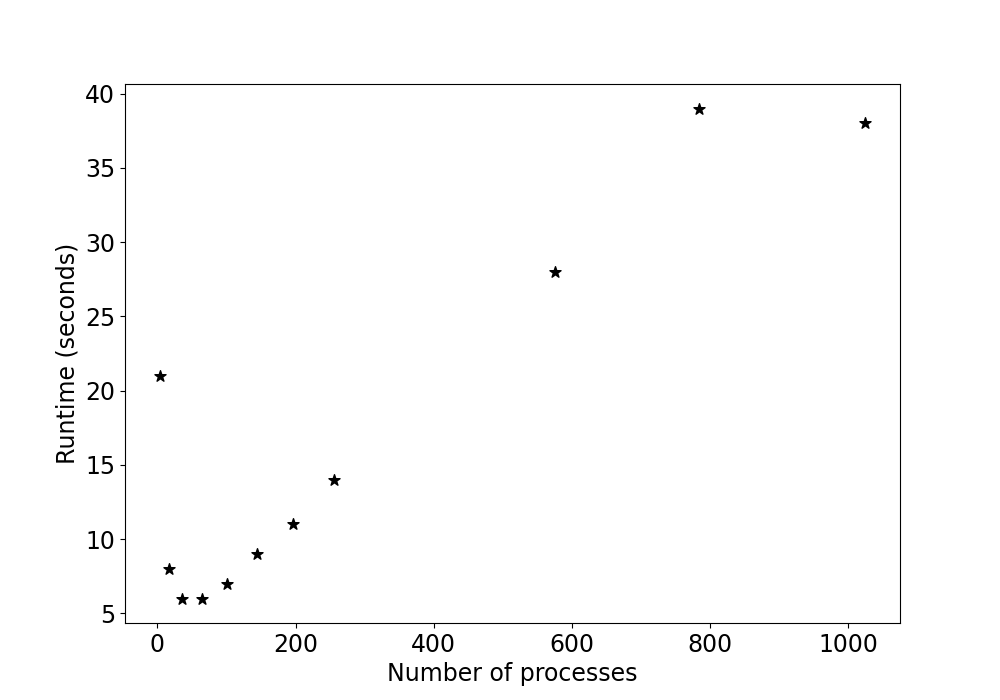}
\caption{Low-Order Strong Scaling of Beatnik on Lassen}
\label{fig:low-strong-scale}
\end{figure}
Figures~\ref{fig:low-weak-scale} and~\ref{fig:low-strong-scale} show the results of these tests. Both cases demonstrate significant scalability problems due to the intense communication demands of large-scale FFTs on modern systems. Specifically, low-order solver runtime increases approximately linearly between 4 and 196 processes and between 256 and 1024 processes but with a smaller slope despite the fact that number of mesh points per GPU is held constant in this case.

Similarly, strong-scaling performance shows a significant performance increase when scaling from 4 to 64 GPUs, a parallel efficiency of only 21\% (3.5x speedup when moving from 4 to 64 GPUS). Similarly, performance turns over and begins to decrease after 64 GPUs due to the small amount of computation and large number of messages that must be sent in this case. In particular, each GPU stores and computes on only a 76 by 76 section of the surface mesh in the 64-node case, but must perform all-to-all communication with dozens of other processes. 

\subsection{High-Order Cutoff Solver Weak Scaling}
\label{subsec:high-weak-scaling}
\begin{figure}[ht!]
\centering
\includegraphics[width=0.7\textwidth]{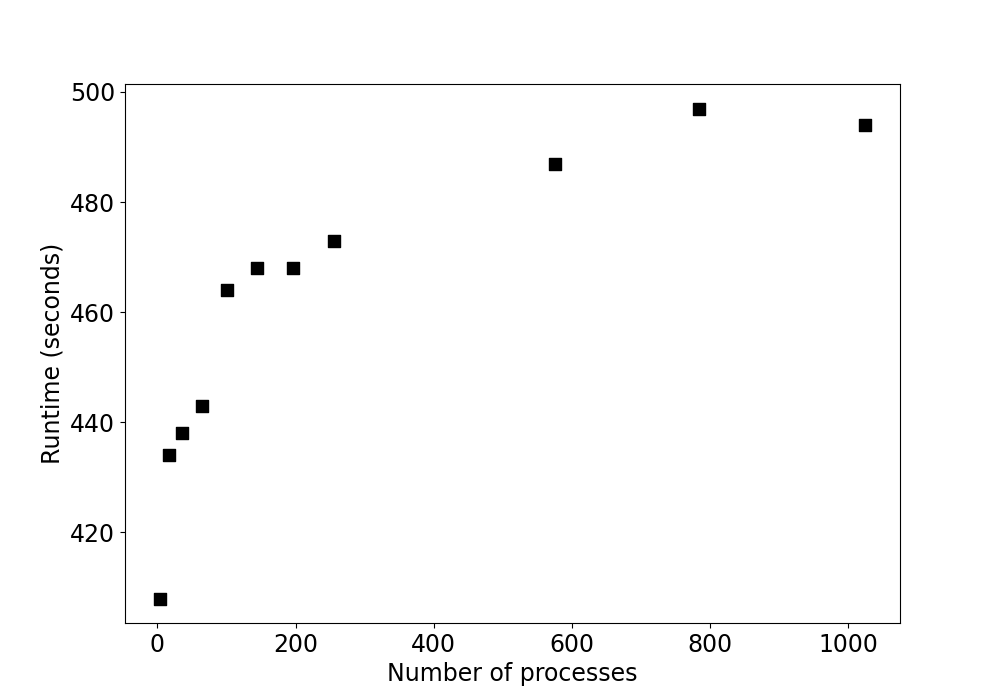}
\caption{Total Runtime When Weak-Scaling the Cutoff Solver}
\label{fig:high-weak-scale}
\end{figure}

To test the scalability of the Beatnik benchmark and evaluate the ability of the underlying communication libraries, we weak scaled the high-order cutoff solver from 4 to 1024 GPUs on Lassen. Figure~\ref{fig:high-weak-scale} shows the results of these tests.  
As expected, weak scaling Beatnik from 4 to 1024 GPUs results in only modest (approximately 20\%) increases in runtime. This is because the amount of computation per GPU remains constant and because communication in the weak-scaling multi-mode case, where little load imbalance develops, is primarily halo exchanges between neighboring ranks in the 2D \texttt{SurfaceMesh} and 3D \texttt{SpatialMesh}. We hypothesize that the modest increase in runtime is due to the overheads of the surface mesh to spatial mesh to surface mesh migration, but again these overheads are generally small, despite a factor of 256 increase in problem size, due to the lack of load imbalance in this test case.

\subsection{Strong Scaling of the Cutoff Solver}
\label{subsec:high-strong-scaling}
As the last scalability evaluation of Beatnik, we tested its ability to create load imbalances between processes, resulting in dynamic and irregular communication. We also evaluated the impact of such communication on be scaling Beatnik from 4 to 256 GPUs on Lassen. As described in Section~\ref{sec:test-cases}, this test uses the single mode problem which is prone to load imbalances as the center of the mesh rolls up as time advances (see Figure~\ref{fig:high-strong-io}).

\begin{figure}[ht!]
\centering
\begin{minipage}{0.45\textwidth}
\includegraphics[width=\textwidth]{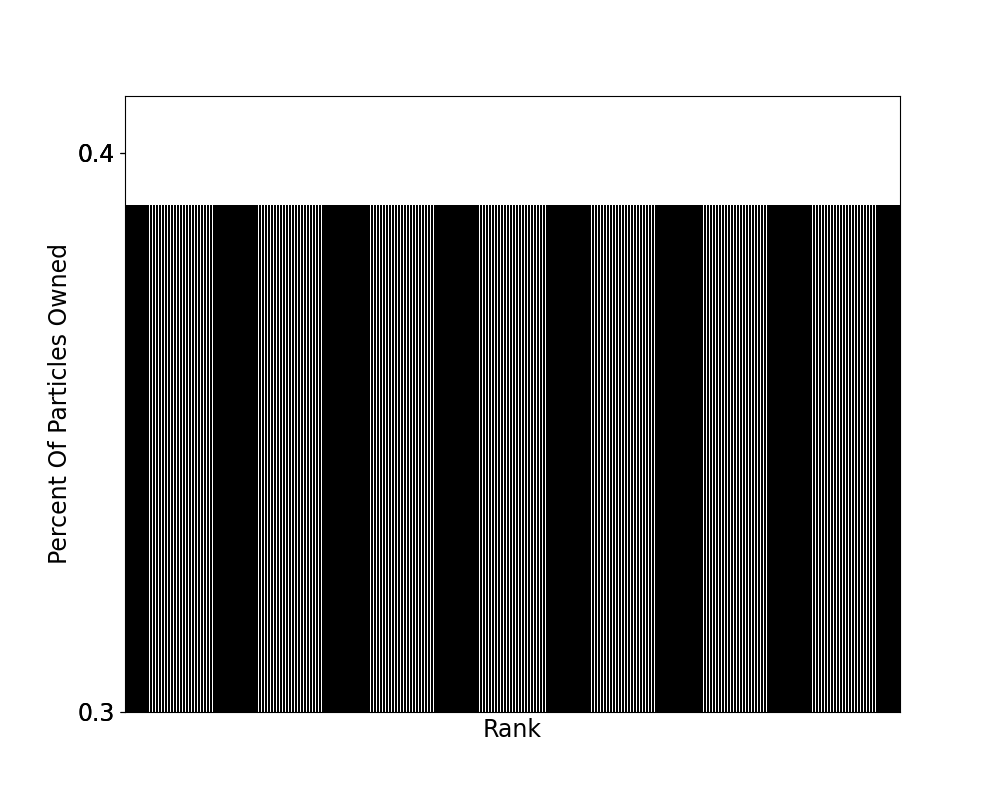}
\caption{Particles Owned by Each of the 256 Ranks at Timestep 80}
\label{fig:high-particles-80}
\end{minipage}
\begin{minipage}{0.45\textwidth}
\centering
\includegraphics[width=\textwidth]{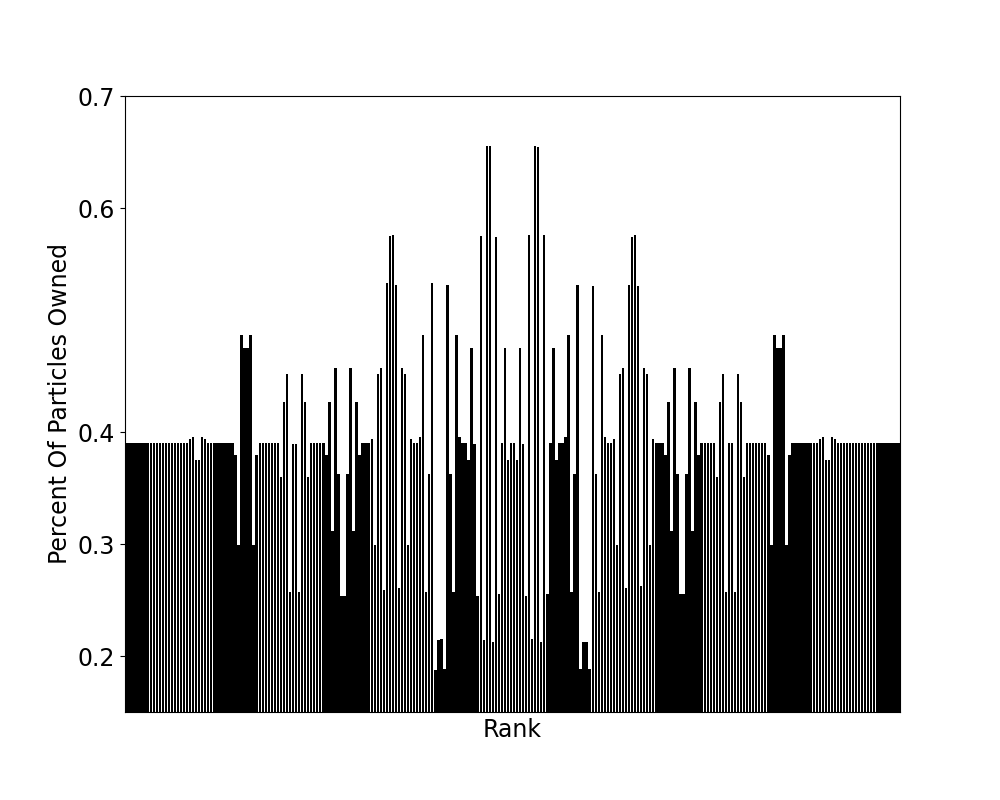}
\caption{Particles Owned by Each of the 256 Ranks at Timestep 340}
\label{fig:high-particles-340}
\end{minipage}
\end{figure}
Figures~\ref{fig:high-particles-80} and~\ref{fig:high-particles-340} show the point distribution amongst 256 processes at timestep 80 and timestep 340. As time advances, the number of particles each processes owns in 3D space becomes increasingly more imbalanced as the interface surface rolls up. At timestep 80, the load is evenly distributed, with all processes owning slightly under 0.4\% of all points. At timestep 340, the load is imbalanced. The processes that own sections of the mesh outside of the rollup have their load stay the same at about 0.4\% of all points. On the other hand, processes within the rollup own between 0.2\% to 0.65\% of all points. 

\begin{figure}[ht!]
\centering
\includegraphics[width=0.7\textwidth]{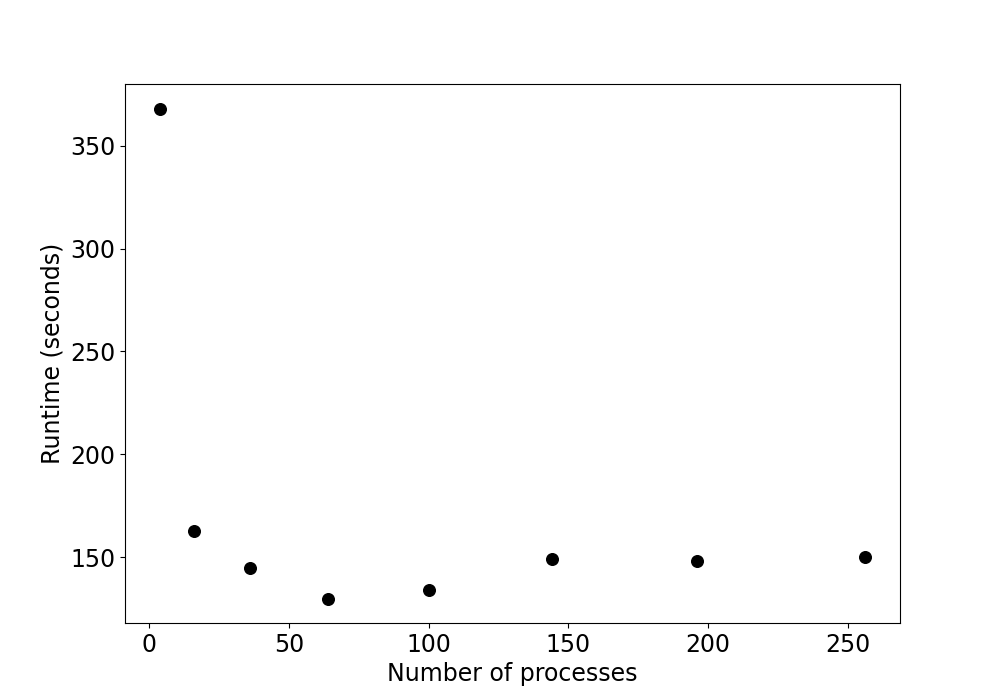}
\caption{Total Runtime of Strong-Scaling the Cutoff Solver}
\label{fig:high-strong-scale}
\end{figure}
Figure~\ref{fig:high-strong-scale} shows the runtime impact of strong scaling on runtime.  Scaling from 4 GPUs to 64 GPUs reduces runtime by factor of 3.3 while the number of GPUs is increased by a factor of 16, a parallel efficiency of 21\%. While performance turns over beyond this point, the performance reduction from additional GPUs is modest because of the localization of communication provided by the cutoff solver.

\subsection{Evaluation of HeFFTe Configurations}
\label{subsec:heffte-evaluations}
Finally, we used the low-order solver in Beatnik to test the performance of different communication configurations of the HeFFTe Fast Fourier Transform library~\cite{krzhizhanovskaya_heffte_2020}. HeFFTe provides three parameters, which can be set to \textit{True} or \textit{False}, to control how Fast Fourier Transform communication and computation is performed: All-to-All, Penciling, and Reordering. 

To do so, we measured low-order solver performanec on all eight permutations of these configurations to demonstrate Beatnik's ability to expose how changes in communication parameters affect benchmark run time; the tested configurations are shown in Table~\ref{table:heffte-configs} and numbered for ease of reference in figures.
\begin{table}[ht!]
    \centering
    \caption{HeFFTe parameter configurations on the low-order solver}
    \label{table:heffte-configs}
    \begin{tabular}{p{0.2\linewidth} p{0.15\linewidth} p{0.15\linewidth} p{0.15\linewidth} p{0.15\linewidth}}
   Configuration		& AllToAll		& Pencils		& Reorder \\

\hline
0 & False & False & False \\
1 & False & False & True \\
2 & False & True & False \\
3 & False & True & True \\
4 & True & False & False \\
5 & True & False & True \\
6 & True & True & False \\
7 & True & True & True \\
\hline

    \end{tabular}
\end{table}

For these tests, we used the base problem size specified in Section~\ref{subsec:setup} weak scaled from 4 to 1024 processes.

\begin{figure}[ht!]
\centering
\includegraphics[width=0.9\textwidth]{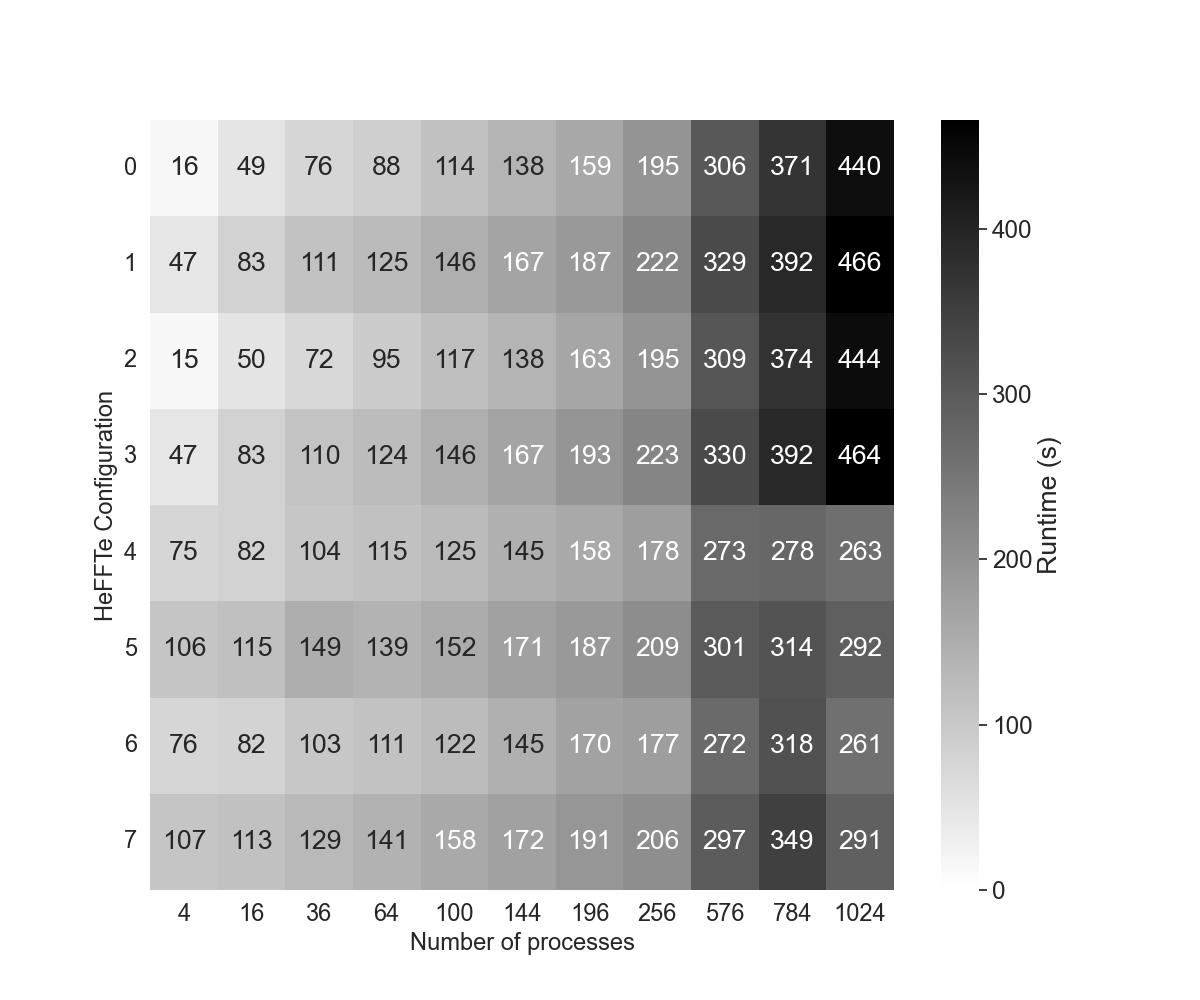}
\caption{Total Runtime When Weak-Scaling HeFFTe Configurations}
\label{fig:heffte-weak-scale}
\end{figure}

Figure~\ref{fig:heffte-weak-scale} show the Beatnik runtimes of each of these configurations shaded by the total runtime.
Beatnik exposes important performance differences between different heFFTe configurations. Most notably, heFFTe performance using the \emph{AllToAll} parameter, which instructs it whether or not to use the builtin \verb|MPI_Alltoall| primitive, varies significantly between small and large numbers of processes. On small numbers of processes, heFFTe performance is better when using its custom communication routines and not using Spectrum MPI's \verb|MPI_Alltoall| primitive. In contrast, on large numbers of processes, heFFTe performance improves if the AllToAll parameter is true. 

\section{Future Work}
\label{sec:future}

We have identified multiple areas of future work in Beatnik. First, Beatnik would be significantly improved by the enhancement of existing and implementation of additional Birchoff-Rott solvers. For the existing cutoff solver, implementation of Cabana sparse mesh support would add additional load balancing communication steps to the benchmark and increase the variety of communication patterns that could be studied. Similarly, the addition of fast multipole and P3M far-field force solvers would also increase Beatnik's capabilities for examining a wide variety of global communication patterns. Second, adding remeshing support to Beatnik, a numerical feature still in development by the Z-model's authors, would allow it to redistribute or add points to the surface mesh as the simulation developed. This would both enable the benchmark to evolve to greater levels of load imbalance and benchmark additional important global communication patterns. Third, we would like to examine both the performance and accuracy of the medium-order model when used with the cutoff solver. Because the medium-order model uses FFTs for calculating changes in vorticity and supports larger timesteps than the high-order model, the performance and accuracy tradeoffs between the two models are potentially interesting. Finally, we would like to enhance Beatnik to support additional test cases and boundary conditions, including more test cases examined by the original Z-Model authors, non-periodic boundary conditions for low- and medium-order solves, and periodic boundary conditions for scalable high-order solves. 

\section{Conclusion}
\label{sec:conclusion}

Beatnik is a novel mini-application designed to support the evaluation of a wide range of complex communication patterns not frequently tested by other mini-applications or benchmarks. Beatnik's implementation is open source and available for download from github~\cite{bridges_beatnik_nodate}, and leverages multiple open-source solver libraries and performance portability libraries to provide to significant computational capabilities in a compact, easy-to-understand code base. As part of Beatnik's development, we have identified multiple test cases, each of which exercises a different, important communication feature in modern HPC systems. Finally, we have demonstrated both the scalability of Beatnik on modern GPU systems and its ability to evaluate the impact of communication parameter changes on benchmark performance.



\begin{credits}
\subsubsection{\ackname}
This work was performed with partial support from the National Science Foundation under Grants Nos.~OAC-2103510, by the U.S. Department of Energy's National Nuclear Security Administration (NNSA) under the Predictive Science Academic Alliance Program (PSAAP-III), Award DE-NA0003966, and by the U.S. Department of Energy through the Los Alamos National Laboratory (LANL). LANL is operated by Triad National Security, LLC, for the National Nuclear Security Administration of U.S. Department of Energy (Contract No. 89233218CNA000001). 

The authors would also like to thank Prof.~Steve Shkoller for providing Z-Model Matlab implementations and test cases for use in the development of the Beatnik benchmark.

\subsubsection{\discintname}
The authors have no competing interests to declare that are relevant to the content of this article.
\end{credits}

\printbibliography

\end{document}